\documentclass[11pt]{iopart}
\usepackage{bm}
\usepackage{hyperref}
\usepackage{amssymb}
\usepackage{feynmp}
\usepackage{amsopn}
\usepackage{slashed}
\RequirePackage{ifpdf}

\ifpdf
\else % ordinary latex seems to include these as eps files without a problem
\fi\usepackage{graphicx}

\newcommand{\Mp}{M_{\mathrm{P}}}
\newcommand{\bo}{\bar{\omega}}
\newcommand{\Bc} {B_{\rm chameleon}}
\begin{document}

	\begin{flushright}
		\textsf{DESY 10-114}
	\end{flushright}
	\title{Chameleon Induced Atomic Afterglow}
	
	\author{Philippe Brax$^{1}$ and  Clare Burrage$^2$ }
	\vspace{3mm}
	
	\address{$^1$ Institut de Physique Th\'{e}orique,
	  CEA, IPhT, CNRS, URA2306, F-91191 Gif-sur-Yvette c\'{e}dex,
	  France \\[3mm]
	$^2$ Theory Group,	 Deutsches Elektronen-Synchrotron DESY,	 D-22603,
	  Hamburg, Germany }

	\vspace{3mm}

	\eads{\mailto{philippe.brax@cea.fr},
	\mailto{clare.burrage@desy.de}}
	
	\begin{abstract}
The chameleon is a scalar field whose mass depends on the density of
its environment.   Chameleons are necessarily coupled to matter particles and will excite transitions between atomic energy levels in an
analogous manner to photons. When created inside an optical cavity by
passing a  laser beam  through a constant magnetic field,  chameleons are trapped
between the cavity walls and form a standing wave.  This effect will lead to an afterglow
phenomenon even when the laser beam and the magnetic field have been
turned off, and could be used to probe
the interactions of the chameleon field with matter.

	\end{abstract}
	\maketitle

	\section{Introduction}
	\label{sec:introduction}

Light scalar fields are commonly invoked as the solution to  various
cosmological problems, including the origins  of dark energy and inflation. They are also generic in theories of beyond the
standard model particle physics.  Problematically, these light scalar fields will also
mediate new fifth forces which have not yet been seen and are very
tightly constrained by experiments \cite{Will:2001mx}.

In \cite{Khoury:2003rn,Khoury:2003aq} Khoury and Weltman proposed a mechanism by which light
scalar fields could evade  gravitational tests of fifth forces.  If the scalar field couples to matter in a non-minimal
way then non-linearities in the theory mean that the mass of the
scalar field becomes dependent on its environment.  By becoming heavy
in dense environments and light in diffuse ones the scalar field
avoids experimental probes of gravity thanks to what is  known as the
{\it thin shell mechanism}.  We shall call scalar fields which behave
in this way {\it chameleons}\footnote{There exist related models of
  scalar fields with environment dependent properties \cite{Luty:2003vm,Nicolis:2008in} that
  avoid the constraints of fifth force experiments \cite{Burrage:2010rs} but do not
  exhibit the afterglow phenomenon discussed in this article.}.

The dynamical way in which chameleonic scalar fields avoid detection
in experiments means that their coupling to matter fields can be as
strong as, or stronger than, the gravitational coupling \cite{Mota:2006fz}.  This
invites us to consider ways in which the chameleon could be detected,
in particular those experiments conducted in near vacuum where the
chameleon cannot hide by adjusting its mass.  Many constraints on
the properties of the chameleon have been obtained in this way by proposing an
additional coupling of the scalar field to photons \cite{Brax:2007ak},
and such a coupling
was  shown to be generic in extensions of the standard model
\cite{Brax:2009ey}.  A coupling between photons and chameleons allows for many
new searches for the chameleon both in the laboratory \cite{Brax:2007ak,Brax:2007hi,Gies:2007su,Ahlers:2007st,Chou:2008gr,Upadhye:2009iv,Brax:2009aw,Brax:2009ey,Rybka:2010ah} and with
astrophysics \cite{Burrage:2007ew,Burrage:2008ii,Burrage:2009mj,Davis:2009vk,Schelpe:2010he,Avgoustidis:2010ju,Levshakov:2010tj}.

Of particular interest for this work are the laboratory searches known
as {\it afterglow} experiments \cite{Gies:2007su,Ahlers:2007st,Chou:2008gr}.  In such experiments a vacuum tube is
placed into a magnetic field, and a laser beam is shone through the
tube.  In the presence of the magnetic field there is a probability $P_{\gamma\leftrightarrow\phi}$
that a photon in the laser light may oscillate into a
chameleon particle \cite{Raffelt:1987im}.  The chameleons produced in this way
cannot exit from the vacuum tube, as they do not have enough energy to
adjust their mass sufficiently to pass through the
wall.  Once the laser source has been turned off we are left with a vacuum
tube full of chameleons.  However if the magnetic field remains, there
is still a probability $P_{\gamma\leftrightarrow\phi}$ that the
chameleon will convert back into a photon.  These reconverted photons
are known as the afterglow and can be detected.  Afterglow searches
for chameleonic fields have been performed with optical
photons by GammeV \cite{Chou:2008gr,Upadhye:2009iv} and with microwave photons by ADMX \cite{Rybka:2010ah}.

If chameleons are produced and their effects detected in afterglow
experiments we will have a good probe of their mass and the strength
of their coupling to photons.  A complete theory of the chameleon also
requires knowledge of  their coupling to matter
fields.  This coupling is currently only weakly constrained by
gravitational experiments \cite{Mota:2006fz} and probes of the  Casimir force \cite{Brax:2007vm}.  In this
article we describe a new experiment to probe the coupling of the
chameleon to matter.

We will show that the chameleon can excite transitions between atomic energy levels, in a similar
manner to the usual photon induced transitions.  However as the
chameleon is a scalar field, whilst the photon is a vector field,
different transitions will be excited by the two fields.  In this
article we focus on the
scenario in which an electron is excited from a lower to a higher
energy level by a chameleon field, and then decays to a lower energy
level by emission of a photon.  The detection of these photons would
allow the coupling of the chameleon to photons to be probed directly.

The  paper is organized as follows. In a first section, we describe the interaction of the chameleons with bound electrons in atoms. We then solve the
two level system when bound electrons hop between two atomic levels
due to the chameleon interaction. We   show that for most experimental situations, the chameleon-matter interaction is weak and that the end result of the chameleon interaction with bound electrons is the creation of a long lived photonic afterglow.

\section{The Chameleon and its Interaction with Matter}

\subsection{Chameleon Models}

The simplest action for a chameleon field, $\phi$, is \cite{Khoury:2003rn,Khoury:2003aq}
\begin{equation}
S=\int d^4x\;
\sqrt{-g}\left\{\frac{M_P^2}{2}R-\frac{1}{2}(\partial\phi)^2-V(\phi)\right\}
- \int d^4x \; \mathcal{L}_m(\psi^{(i)}, \tilde{g}_{\mu\nu})\;,
\label{eq:action}
\end{equation}
where the matter fields $\psi^{(i)}$ feel the Jordan frame metric
$\tilde{g}$ which is related to the Einstein frame metric $g$ by
\begin{equation}
\tilde{g}_{\mu\nu}=e^{2\beta\phi}g_{\mu\nu}\;.
\label{eq:coupling}
\end{equation}
  $V(\phi)$ is an arbitrary scalar
potential.  The Einstein and Jordan frame descriptions are completely
equivalent classically.  In the Einstein picture the masses and couplings of elementary particles become
dependent on the chameleon field.  With such a conformal coupling
(\ref{eq:coupling})  there is no tree level
interaction between the scalar field and photons, however we are
allowed to add such a term to the action
\begin{equation}
S_{\gamma}=\int d^4x\;\sqrt{-g}\phi\beta_{\gamma}F_{\mu\nu}F^{\mu\nu}\;,
\end{equation}
the existence of which follows from quantum effects \cite{Brax:2009ey}.

When the equations of motion for the fields are computed from the
action (\ref{eq:action}), it becomes apparent that the dynamics of the chameleon are
governed by an effective potential
\begin{equation}
V_{eff}(\phi)=V(\phi)-e^{4\beta\phi}\tilde{T}^{\mu}_{\mu}\;,
\end{equation}
where $\tilde{T}^{\mu}_{\nu}$ is the usual Jordan frame energy
momentum tensor for matter fields. We restrict our attention to the
behaviour of the chameleon in the presence of  non-relativistic matter
for which
$\tilde{T}^{\mu}_{\mu}=-\rho$, and  $\rho$ is the energy density.

For the chameleon mechanism to function the bare potential, $V(\phi)$, must
be chosen so that the effective potential has a minimum,
$V_{eff}^{\prime}(\phi_m)=0$, and the position of this minimum  is a function of the local energy density. The chameleon
mechanism also requires  that the mass of small fluctuations about
this minimum
\begin{equation}
m^2 =V^{\prime\prime}(\phi_m)+\beta^2\rho e^{4\beta\phi}\;,
\end{equation}
is a monotonic function of the local energy density, and that the mass of the chameleon field increases with increasing density.

The most stringent constraints on $\beta$ come from measurements of
the Casimir force \cite{Brax:2007vm}.  For potentials of the form
$V=\Lambda_0^4[1+(\Lambda/\phi)^n]$ where $|n|\sim\mathcal{O}(1)$ and
$\Lambda_0=\Lambda=2.4\times 10^3\mbox{ eV}$ the experiments constrain
$\beta\mbox{ GeV}\gtrsim10^{-16}$.  For other forms of the potential the constraints on $\beta$
are less stringent.  Constraints from particle physics experiments,
\cite{Brax:2009aw}, give $\beta\mbox{ GeV}
\lesssim 10^{-4}$.

\subsection{Chameleon-Atom Interactions}
In this article we describe a new technique to probe the coupling of
the chameleon to matter fields if they are produced in afterglow
experiments.  The coupling of the chameleon to the matter fields in
the action  (\ref{eq:action}),
 implies an interaction between a background chameleon field
and electrons residing in atoms. The existence of such a coupling means that a
chameleon field
can excite atomic energy level transitions, in the same way that a
background photon field can give rise to atomic absorption and emission. To study how the chameleon
field can excite electrons and induce transitions between energy levels we
begin by deriving the perturbed Hamiltonian governing the behaviour of
the electrons in the presence of a chameleon field.

The Dirac equation for a two component spinor is
\begin{equation}
i\frac{\partial}{\partial t}\left(\begin{array}{c}
\chi_+\\
\chi_-
\end{array}\right)=\left(\begin{array}{cc}
m(\phi) & \sigma\cdot p\\
\sigma\cdot p & -m(\phi)
\end{array}\right)\left(\begin{array}{c}
\chi_+\\
\chi_-
\end{array}\right)\;,
\end{equation}
where the mass of the spinor is dependent on the value of the
chameleon field.
Under the transformation
\begin{equation}
\left(\begin{array}{c}
\chi_+\\
\chi_-
\end{array}\right)=\exp\left(-i\int m\; dt\right)\left(\begin{array}{c}
\Phi\\
X
\end{array}\right)\;,
\end{equation}
 the equations of motion become
\begin{eqnarray}
i\dot{\Phi}&=& e^{i\int m\; dt}\sigma\cdot p (e^{-i\int m\; dt} X)\;,\\
i\dot{X}&=& e^{i\int m\; dt}\sigma\cdot p (e^{-i\int m\; dt} \Phi) -
2m X\;,
\end{eqnarray}
and we impose that $X$ does not vary
significantly with time:
\begin{equation}
X=\frac{1}{2m}e^{i\int m\; dt}\sigma\cdot p (e^{-i\int m\; dt} \Phi)\;.
\end{equation}
A  final transformation  $\psi =e^{-i\int m\; dt} \Phi$ gives the Schrodinger-type equation
\begin{equation}
i\dot{\psi}=\left[\frac{p^2}{2m}+m(\phi)+\frac{1}{2}(\sigma\cdot
  p)\left(\frac{1}{m(\phi)}\right)(\sigma\cdot p)\right]\psi\;.
\end{equation}

Assuming that the variation of $\phi$ about its background value $\phi_0$
is small we can approximate the mass of the electron $m(\phi)=m_e(1+\beta \delta\phi)$.
 Then to
first order in the chameleon fluctuation the Hamiltonian becomes
\begin{equation}
H=\frac{p^2}{2m_e}+m_e-\frac{\beta}{2m_e}\left[\delta\phi
  p^2+(\sigma\cdot p) \delta\phi (\sigma\cdot p)\right]  +m_e\beta\delta\phi\;.
\label{Ham}
\end{equation}

\subsection{Transitions Between Energy Levels}

The chameleon perturbation of the electron Hamiltonian, derived in the previous
section, allows electrons to have transitions from one energy level of an
atom to another.  If $|i\rangle$ is the i-th excited state of an atom
the transition rate between one energy level and another due to the
chameleon is $\langle i | \delta H(\phi)|j\rangle$, where
\begin{equation}
\delta H=-\frac{\beta}{2m_e}\left[\phi
  p^2+(\sigma\cdot p) \phi (\sigma\cdot p)\right]  +m_e\beta\phi\;,
\end{equation}
and we now denote the chameleon fluctuations by $\phi$.

In a vacuum tube, aligned along the $z$ axis, that has been filled with
chameleons by conversion from photons with frequency $\bo$,  the chameleon field can be
written as
\begin{equation}
\phi(t,z)=-a\cos\bo t\cos\bo z\;,
\end{equation}
where $a$ is a constant that will be determined in Section \ref{sec:cavity}.

The wave-functions describing the energy levels of an hydrogenic atom
are exponentially suppressed outside the Bohr radius of the atom, $a_0$.
Therefore it is sufficient to expand the chameleon wave function about
the position of an atom, $z_0$, on the $z$-axis
\begin{equation}
\phi(t,z_0+r\cos\theta)=-a\cos\bar{\omega} t\times(\cos\bar{\omega} z_0-\bar{\omega}
r \cos\theta\sin\bar{\omega} z_0)\;.
\label{pertphi}
\end{equation}
Where $r$, $\theta$ and $\phi$ are spherical polar coordinates
centered at the nucleus of the atom.  We have assumed $r$ is small
compared to the scale of variation in the chameleon wavefunction.

We consider transitions between the 1s and 2p energy levels of a
hydrogenic atom, with nuclear charge $Ze$ \footnote{The chameleon can also induce transitions from the 1s to the 2s level. This transition is
optically forbidden as optical transitions are mostly due to a dipole interaction which changes the parity of the wave functions.}.
The wave functions for these states are (in spherical polar coordinates)
\begin{eqnarray}
\psi_{1s}&=&\frac{1}{\sqrt{\pi}}\left(\frac{Z}{a_0}\right)^{3/2}e^{-Zr/a_0}\;,\\
\psi_{2p}&=&\frac{1}{\sqrt{\pi}}\left(\frac{Z}{2a_0}\right)^{5/2}e^{-Zr/2a_0}r\cos\theta\;.
\end{eqnarray}
An order of magnitude estimate for the size of  $\langle i | \delta
H(\phi)|j\rangle$, with $H(\phi)$ given in (\ref{Ham}), shows that the
term due to  $m_e\beta\phi$ will
always dominate over those that are a function of the momenta. The
transition probabilities due to this term in the Hamiltonian are
\begin{equation}
\langle
1,0|m_e\beta\phi|2,1\rangle=\frac{2^8\pi }{\sqrt{2}3^5Z}m_e\beta
  a_0\bar{\omega}a\;\sin\bar{\omega}
z_0 \cos\bar{\omega} t\;,
\end{equation}

\begin{equation}
\langle
1,0|m_e\beta\phi|1,0\rangle=\langle
2,1|m_e\beta\phi|2,1\rangle=-m_e\beta a \cos\bar{\omega}
z_0\cos\bar{\omega} t\;.
\end{equation}

\subsection{Chameleons in a Cavity}
\label{sec:cavity}
To fully calculate the transition probabilities induced by the
chameleon we need to know the wavefunction describing the behaviour of
chameleons in a cavity.  We envisage a scenario where the cavity has
been filled with a bath of chameleons by an afterglow-like experiment.  A simplified form
of the set up has the photons entering the tube at $z=0$ and passing
straight through to exit
at $z=L$.  The chameleons are reflected at $z=0,L$.

Following the derivation of \cite{Gies:2007su} and assuming that, as we are
dealing with laser sources, the incoming
photon beam is very sharply peaked about the frequency $\omega=\bo$,
we find the following expression for the chameleon field inside the cavity
\begin{eqnarray}
\phi(t,z)&=&-i\vartheta e^{-i\bar{\omega}(t-z)}\alpha(\bar{\omega})\label{eq:wave1}\\
& &\times\left(1+\frac{\bar{\omega}e^{-i\bar{\omega}z}}{k_+\sin
    k_+L}(\sin k_+(z-L)+e^{i\bar{\omega} L}\sin
  k_+z)\right)+\mathcal{O}(\vartheta^2)\;,\nonumber
\end{eqnarray}
where $\vartheta$ is the angle describing the mixing between
chameleons and photons, which is assumed to be small; $\vartheta
\approx \bo B\beta_{\gamma}/m^2\Mp$ \cite{Raffelt:1987im}.  Here $m$ is the mass of the
chameleon inside the cavity and $B$ the strength of the magnetic field.  $\alpha(\omega)$ is the amplitude of
oscillations of  the
incoming photons, and $k_+=\bo-m^2/2\bo$.

Once a chameleon field is present in the cavity the magnetic field is
turned off at time $t=0$.  The chameleons now
propagate freely inside the cavity.  The solution for the chameleon
wavefunction at time $t>0$ can be found by Fourier expanding the
wavefunction and matching with (\ref{eq:wave1}) at time $t=0$.  The standing
wave solution is
\begin{equation}
\phi=-2\vartheta\bo L\alpha(\bo)\sin\bo L\sum_{n\geq
  0}\frac{\cos \omega_n t\cos k_n z}{\pi^2n^2-k_+^2L^2}(1+(-1)^n\cos k_+L)\;,
\end{equation}
where $k_nL=n\pi$ and $k_n^2=\omega_n^2-m^2$.
As $k_+ = \bar{\omega}(1+\mathcal{O}(m^2/\bo^2))$ and we assume that
$m\ll\bo$, as is the case for the GammeV experiment.  The dominant
contributions are the terms with $\vert n\vert \sim \bo L$.  Assuming
that the length of the cavity is not tuned to be resonant with the optical frequency 
$\bo^2L^2-n^2\pi^2$ never vanishes, and the leading order behaviour of the
chameleon wavefunction is
\begin{equation}
\phi(t,z)=-a\cos\bar{\omega} t \cos\bar{\omega} z\;,
\end{equation}
where $a=a_-+a_+$ together with
\begin{equation}
a_\pm=\frac{4\vartheta
  \bar{\omega}L\alpha(\bar{\omega})}{(\bar{\omega}^2L^2-N^2\pi^2)}\sin\bar{\omega}L(1\mp\cos\bo L)\;,
\end{equation}
and $N$ is the closest integer to $\bo L/\pi$.
\section{The Two State System}

\subsection{Evolution of the Electron Bound State}

We study a simplified system in which an atom has only two
energy levels.  This  is a good approximation to the excitation of the
ground state of an atom when there is a unique gap between levels that
is close to the energy of the chameleon field, however we do not require an
exact resonance.  The approximation of
a two state system breaks down if the number of excited states with an energy gap
close to the chameleon energy is greater than one.   We label the lower and
upper levels 1 and 2 respectively.  The electron wavefunction is described by $|\psi \rangle
=c_1|1\rangle +c_2|2\rangle$, where $c_1$ and $c_2$ are the
probability amplitudes that the electron will be found in the first
and second energy level respectively.  For full discussion of photon
induced transitions in a two state system see, for example, \cite{Fox}.

The coefficients $c_i$ evolve with time according to
\begin{equation}
\frac{\partial}{\partial t}\left(\begin{array}{c}
c_1\\
c_2
\end{array}\right)=ia\cos\bar{\omega} t\left(\begin{array}{cc}
A & -B e^{-i\omega_0t}\\
-B e^{i\omega_0t} & A
\end{array}\right)\left(\begin{array}{c}
c_1\\
c_2
\end{array}\right)\;,
\label{matrix}
\end{equation}
where $\omega_0$ is the energy difference between the two energy
levels under consideration, for  two energy levels of a
hydrogenic atom  $\omega_0\sim{\cal O}(\mbox{eV})$.  We have also defined
\begin{eqnarray}
A &=& m_e\beta\cos\bar{\omega} z_0\;,\\
B &=& \frac{2^8m_e\beta
  a_0\bar{\omega}\pi}{\sqrt{2}3^5Z}\sin\bar{\omega} z_0\;.
\end{eqnarray}
Writing
\begin{eqnarray}
c_1&=&\tilde{c}_1\exp\left(-\frac{iA
    a}{\bar{\omega}}\sin\bar{\omega} t\right)\;,\\
c_2&=&\tilde{c}_2\exp\left( -\frac{iA
    a}{\bar{\omega}}\sin\bar{\omega} t \right)\;,
\end{eqnarray}
 the equations can be diagonalised
\begin{equation}
\frac{\partial}{\partial t}\left(
\begin{array}{c}
\tilde{c}_1\\
\tilde{c}_2
\end{array}\right) =-iaB\cos\bar{\omega} t \left(\begin{array}{cc}
0 & e^{-i\omega_0 t}\\
e^{i\omega_0 t} & 0
\end{array}\right)\left(
\begin{array}{c}
\tilde{c}_1\\
\tilde{c}_2
\end{array}\right)\;.
\end{equation}
As is the case for photon driven excitations, there are two limits in which these
equations can be solved analytically, called the weak and strong
field limits.  We discuss these in detail in the following sections.

\subsection{The Weak Field Limit}

If the electron-chameleon coupling is weak then the number of atoms in the
lowest energy level is always much greater than the number in the
excited energy level.  The approximations
$\tilde{c}_2\ll\tilde{c}_1\approx 1$,  valid when $\vert A a\vert \ll \bar \omega $, reduce the system of equations to
\begin{equation}
\frac{\partial \tilde{c}_2}{\partial t} = iaBe^{i\omega_0
  t}\cos\bar{\omega} t\;.
\end{equation}
To integrate this equation we must recall that  both the
spectral line and the laser beam have a finite spectral width; in the
evaluation of $a$ there is an integral over frequency which we have
assumed to be infinitely sharply peaked at $\omega=\bo$. Replacing
$a^2= \int u(\omega)d\omega$ allows us to
integrate over the spectrum of the laser beam, where we choose the laser profile to be approximated with a Lorentzian distribution of width $\Delta \omega$,
\begin{equation}
u(\omega)=\frac{a^2 \Delta \omega}{\pi} \frac{1}{
  (\omega-\bar\omega)^2 +\Delta\omega^2}\;.
\end{equation}
The spectral line also has a Lorentzian distribution centered on the
frequency $\omega_0$ with width  $\Delta\omega_0$.
We deduce that:
\begin{equation}
|\tilde{c}_2|^2=B^2\int^{\omega_0+\Delta\omega_0/2}_{\omega_0-\Delta\omega_0/2}u(\omega)\left(\frac{\sin(\omega-\omega_0)t/2}{\omega-\omega_0}\right)^2\;
d\omega\;. 
\end{equation}
We
assume that $u(\omega)$ does
not vary over the width of the spectral line, which is valid either
when the laser is tuned to the resonance $\omega_0=\bo$ and
$\Delta\omega>\Delta\omega_0$, or when the spectral line
occurs in the tail of the laser spectrum $|\bo-\omega_0|\gg
\Delta\omega$.    Then we find 
\begin{equation}
|c_2(t)|^2=\frac{B^2 \pi}{2}u(\omega_0)t\;.
\end{equation}
after imposing the initial condition $c_2(0)=0$.  The chameleon Einstein coefficient is
\begin{equation}
B_{\rm chameleon}= u(\omega_0) \frac{\pi \langle B^2\rangle}{2}\;,
\label{Bcham}
\end{equation}
where we have averaged over the position of the atom in the cavity
\begin{equation}
\sqrt{\langle B^2\rangle}=\frac{2^7m_e\beta
  a_0\bar{\omega}\pi}{3^5}\;.
\end{equation}

The excited level can be populated thanks to the chameleons and depleted due to photons. Therefore
\begin{equation}
\frac{dN_2}{dt}= B_{\rm chameleon} N_1 -B_{\gamma} N_2\;,
\end{equation}
due to the emission of photons, where
$B_{\gamma}=(\pi/3\epsilon_0\hbar^2)\mu_{12}^2$ is the photonic
Einstein coefficient and $\mu_{12}$ is the dipole matrix element.   The photons leave the cavity very
rapidly so we neglect photon stimulated emission. 
Similarly, the number of electrons in the ground state evolves according to
\begin{equation}
\frac{dN_1}{dt}= -B_{\rm chameleon} N_1\;,
\end{equation}
as chameleons are the only particles permanently present in the cavity.
Finally the number of emitted photons is
\begin{equation}
\frac{dN_{\gamma}}{dt}= B_\gamma N_2\;,
\end{equation}
We find that
\begin{equation}
N_1= N_1(0) e^{-B_{\rm chameleon} t}\;.
\end{equation}
We reach a steady state as long as $t\ll 1/B_{\rm chameleon}$ in which
\begin{equation}
N_2= \frac{B_{\rm chameleon}}{B_\gamma} N_1\;,
\end{equation}
and the number of created photons per unit time is constant
\begin{equation}
\frac{dN_{\gamma}}{dt}= B_{\rm chameleon} N_1\;.
\end{equation}
Hence $B_{\rm chameleon}$ gives the number of photons per unit time emitted by a single atom due to the interaction with chameleons.
It depends on $u(\omega_0)$ crucially implying that if $\omega$ and
$\omega_0$ are not close, the density $u(\omega_0)$ is going to be
tiny. On the other hand if there is a quasi-resonance then we may
create many photons.

\subsection{The Strong Field Limit}
In the limit of an exact resonance we can solve the equations of
motion without  restricting our
attention to the situation where the perturbations due to the
chameleon field are small.  We again make a rotating wave
approximation so that the two state system is described by
\begin{equation}
\frac{\partial}{\partial t}\left(
\begin{array}{c}
\tilde{c}_1\\
\tilde{c}_2
\end{array}\right) =-\frac{iaB}{2} \left(\begin{array}{cc}
0 & 1\\
1 & 0
\end{array}\right)\left(
\begin{array}{c}
\tilde{c}_1\\
\tilde{c}_2
\end{array}\right)\;,
\end{equation}
with solution
\begin{eqnarray}
\tilde{c}_1(t)&=&\cos(\Omega_Rt/2)\;,\\
\tilde{c}_2(t)&=&i\sin(\Omega_R t/2)\;.
\end{eqnarray}
Where
\begin{eqnarray}
\Omega_R&=&|-aB|\\
&=& \frac{2^{10}m_e\beta a_0\bar{\omega}^2 L
  \alpha(\bar{\omega})\vartheta}{\sqrt{2}3^5Z(\bar{\omega}^2L^2-N^2)}\sin\bo L\sin\bo z_0(1\mp \cos\bo\pi L)\;,
\end{eqnarray}
is known as the Rabi frequency.
The probability for finding the electron in the upper, or lower, level
oscillates, a behaviour known as Rabi oscillations.

\subsection{Predictions for a GammeV-like Afterglow Experiment}
To see if either of the effects described above are detectable with
current experimental set ups we specialize to the specifications of a
simplified form of the Fermilab GammeV experiment.  
  The GammeV experiment has a cavity of
length $L=6$ m, and a laser beam of power of 160 mJ which operates
 at a frequency of 2.33 eV. The laser beam emits 5 ns wide pulses, and
we assume a beam diameter of 5 mm.  This
implies $\alpha_0=1.2\times 10^{-8}\mbox{ GeV}$.  We assume that the
 the atoms under
consideration inside the cavity have $Z\sim 1$ and transition
frequency  $\omega_0 \sim 1\mbox{ eV}$.  The Bohr radius of  such
hydrogenic atoms  is $a_0=5.3\times
10^{-11}\mbox{ m}$.  We assume that the cavity is not resonant with
the optical frequency so that $\bo^2L^2-N^2\pi^2 \sim
\mathcal{O}(10^{-1})$.

We focus first on the possibility that observable  Rabi oscillations
occur.
The Rabi frequency  depends on the strength of the chameleon
to matter coupling $\beta$ and the photon-chameleon oscillation angle
$\vartheta$
\begin{equation}
\Omega_R\sim10^{19}\vartheta(\beta\mbox{ GeV})\mbox{ s}^{-1}\;,
\end{equation}
assuming the trigonometric terms in $a$ and $B$ are of order one.
The GammeV experiment is sensitive to $\vartheta\gtrsim 10^{-9}$ so in
the strong field limit of a GammeV-like experiment
the chameleon induced Rabi frequency  satisfies $\Omega_R\gtrsim 10^{10}(\beta \mbox{
  GeV})\mbox{ s}^{-1}$.

Rabi oscillations are observable if the period of the oscillations is
less than the radiative life time of the excited energy level.  For a
hydrogen atom, the radiative lifetime of the 2p energy level is
$1.6\mbox{ ns}$.  For more general atoms, we still consider that the lifetime is in the ns range. Hence the Rabi oscillations are only observable if
\begin{equation}
10^{-1}\lesssim \beta\mbox{ GeV}\;,
\end{equation}
which is already excluded by particle experiments.
However it might be possible to observe chameleon driven Rabi
oscillations if the power of the laser were to be increased.

For currently achievable experimental setups, we conclude that the
population of the excited energy level is never significant, and the
system is well approximated by the weak field limit.  For a GammeV-like experiment the number
of photons created per unit time and per atom, $B_{\rm chameleon}$, defined in
(\ref{Bcham}), is
\begin{eqnarray}
\Bc&\approx& \frac{a^2m_e^2a_0^2\beta^2\bo^2\Delta\omega}{(\omega_0-\bo)^2+\Delta\omega^2}\nonumber\\
&\approx& \frac{10^{15}\vartheta^2(\beta\mbox{
    GeV})^2s^{-1}}{\left(\frac{\omega_0}{\mbox{eV}}-1\right)^2+10^{-14}}
\end{eqnarray}
Where we have taken typical values for the laser spectrum with
$\bo\sim 1\mbox{ eV}$ and a width $\Delta\omega \sim 10^{-7}\mbox{
  eV}$. If the laser is not tuned to the resonance $\omega_0/\mbox{eV}-1
\sim\mathcal{O}(\mbox{eV})$.  This implies
\begin{equation}
\Bc\approx{10^{15}\vartheta^2(\beta\mbox{
    GeV})^2s^{-1}}\;.
\end{equation}
Recalling that the GammeV experiment is sensitive only to $\vartheta\gtrsim
10^{-9}$, we see that for a typical experiment of the GammeV type and experimentally allowed values of $\beta$ the
production of photons can be significant:
\begin{equation}
\Bc\gtrsim {10^{-3}(\beta\mbox{
    GeV})^2s^{-1}}\;.
\end{equation}
The typical lifetime of the afterglow phenomenon $1/\Bc$ can be very
large when $\beta$ is small, i.e. macroscopic and of a few seconds. In
this case, the number of created photons can still be relatively large
as the number of atoms in the cavity can be significant.  Again
considering the GammeV experiment which has a length of 6 m and a
diameter of 3.175 cm, and within which the pressure is $\sim
1.9\times10^{-3}\mbox{ torr}$, we find that the number of atoms
within the experiment at room temperature is $\sim 2\times
10^{18}$. The surface area of the aperture compared to the surface
area of the experiment is 0.005, so that we find that emitted photons
should be detected at a rate satisfying
\begin{equation}
\mbox{Rate of detected photons}\gtrsim 10^{13}(\beta\mbox{ GeV})^2s^{-1}\;.
\end{equation}
For sufficiently strong couplings afterglow
photons are easily detectable.

Recently the GammeV collaboration reported \cite{wester} that they have seen an
orange afterglow, the magnitude of which is independent of the
strength of the magnetic field.  Whilst this result is preliminary and
may well be a systematic of the experiment, it is interesting to
speculate, in light of the analysis of this article, that it could be
the atomic emission and absorption caused by a bath of chameleons produced
by the experiment.

\section{Conclusions}
The chameleon is a scalar field with environmental dependent
properties.  Due to a non-linear potential and a non-minimal coupling
to matter the mass of the field is large in dense environments and
small in diffuse ones.  Such a field can have strong couplings to
matter whilst still avoiding the constraints of experimental probes of
gravity.

In such theories an additional coupling of the scalar field to the
photon is common, which enables us to probe the chameleon model in
optical experiments.  The classic laboratory search for such a
chameleon is an afterglow experiment where a laser is shone through a
vacuum tube with a pervading magnetic field. In such an
environment the photons may oscillate into chameleons, these
chameleons will remain trapped in the vacuum tube.  Later, after
the laser has been switched off, the chameleons may oscillate back into
photons.  These photons are known as the afterglow.

Current afterglow experiments are able to probe the mass of the
chameleon and its coupling to photons.  In this article we have
described how they can also be used to probe the coupling of the
chameleon to matter fields.  The coupling to matter means that a bath
of chameleons, inside the vacuum tube of an afterglow experiment,
would excite energy level transitions in atoms present in the vacuum
tube.  This is precisely analogous to atomic emission and absorption
due to photons.  If an electron is excited to a higher atomic energy
level by the chameleon bath, the atom may still decay to a lower
energy level by the emission of a photon.  We have computed the rate
at which such photons would be emitted in a typical afterglow
experiment.  We find that this rate  can be significant, and so searches for this phenomenon can
be used to constrain the chameleon to matter coupling.

Our results are particularly interesting in light of a recent
detection of an orange afterglow in the GammeV experiment, which is
independent of the strength of the magnetic field.  However systematic
explanations for this detection are yet to be excluded.

 All in all, we find that the experimental detection of an afterglow
 phenomenon when the laser beam and the magnetic field have been
 turned off would lead to stringent bounds on a combination of both
 the coupling of chameleons to photons and matter. Depending on the
 values of these couplings, this  afterglow phenomenon could well be a smoking gun for the possible existence of chameleons.
\section*{Acknowledgments}
We would like to thank Axel Lindner, Amanda Weltman and William Wester for very helpful
discussions.  CB is supported by the German Science Foundation (DFG) under the
Collaborative Research Centre (SFB) 676 and would like to thank the Institut de Physique Th\'{e}orique,
	  CEA, for their hospitality while part of this work was completed.

\section*{References}
\bibliographystyle{JHEP}
	\bibliography{electron}
\end{document}